\documentclass{article}
\usepackage{graphicx} 
\usepackage{amsfonts}
\usepackage{amsmath}
\usepackage[
backend=biber,
style=apa
]{biblatex}
\usepackage{booktabs}
\usepackage{float}
\usepackage{appendix}
\usepackage{makecell}
\usepackage{authblk}
\usepackage{hyperref}
\usepackage{authblk}

\title{Beyond the Flow: A Bayesian Latent Clustering Framework for Shared Micro-mobility Users in Venice}

\author[1]{Vanshika Keshwani}
\author[1]{Stefano Mazzuco}

\affil[1]{Department of Statistical Sciences, University of Padova, Italy}

\date{}

\begin{document}

\maketitle



\begin{abstract}
The study on shared micro-mobility is based on trip modeling and user data. User segmentation in shared micromobility systems is traditionally studied by aggregating trip-level observations into user-specific summary measures before applying clustering techniques. Such aggregation can obscure trip-level variability and lead to ecological fallacies if results are interpreted as applying to individual records. We propose a Bayesian finite mixture model for multivariate categorical count data that clusters users directly from repeated trip-level observations while preserving the full categorical structure of individual travel behavior. This approach focuses on identifying heterogeneous mobility users from high-dimensional categorical trip behavior while accounting for uncertainty in cluster assignments. Users are the fundamental unit of analysis for exploring latent cluster patterns. The model represents each user with a product-multinomial likelihood with latent cluster membership.
The methodology is illustrated using a one-year trip record of shared bikes and e-bikes from the Municipality of Venice, Italy, comprising over 220,000 trips made by more than 11,000 recurrent users. The analysis identifies eight distinct latent mobility profiles corresponding to localized, commuter-oriented, tourist-oriented, central, and inter-zonal travel behaviors. The proposed framework provides a flexible and computationally scalable approach for clustering repeated categorical observations and is readily applicable to other large-scale behavioral and transportation datasets.
\end{abstract}


\section{Introduction}
Shared micromobility services have been flourishing in recent years, providing a sustainable and convenient solution in the shared transportation industry. They offer an easy rental service for users, enabling flexibility in the city and filling the gap left by owning a private vehicle. As city administrative departments increasingly invest in shared mobility infrastructures, focusing on the use of these services is essential to optimize efficiency, accessibility, and long-term adoption. 
User acceptance plays an integral role in making any new mode of transportation service a successful model \parencite{samadzadWhatAreFactors2023}. Service development targeting users' behavior can help build a more inclusive environment, increasing the number of users adopting shared bikes. Therefore, micro mobility providers need to identify different user segments, along with their defining characteristics and usage patterns, to tailor operations, infrastructure, and policies effectively. Clustering users aims to create groups of usage patterns from the set of available categorical variables.

The existing literature \parencite{talavera-garciaExaminingSpatiotemporalMobility2021, liSEASONALANALYSISFACTORS2017} on shared micromobility services has focused extensively on studying trip-level characteristics such as distance, duration, and seasonal variations by aggregating riding patterns in shared micromobility services. Although research on shared bike usage provides valuable information, it is limited in its ability to examine user profiles and segment user behavior. Exploring the demand-side perspective and user heterogeneity is essential to understand \textit{who} uses these services and \textit{how} their mobility behavior differs. In this context, clustering users based on their mobility behavior presents a data-driven approach to explore latent user segments.

Considering that we usually have data on two levels (trips are first-level units and users are second-level units), the most trivial way to cluster users is to aggregate trip-level observations by user, deriving measures such as trip frequency, mode-specific origins and destinations, average trip distance, and related variables, and then clustering users based on these aggregated features.  However, while useful at the aggregate level, this approach can obscure trip-level variability and lead to ecological fallacies if results are interpreted as applying to individual records. The present study advances beyond such methods by clustering users while preserving trip-level information, thereby yielding a more accurate representation of user behavior. By identifying the mobility pattern of each user, similarities and dissimilarities can be established among groups, and the groups they fall into represent the important zones, preferred durations, and seasons of their travel. In practice, this is done by adopting a finite mixture model for multivariate categorical data with latent class models, an approach widely used in text mining, genetics, and behavioral profiling.

This model is applied to data of shared mobility services provided in the Municipality of Venice, Italy. The city covers 414.6 km2, with a complex urban structure and a mix of resident and tourist mobility demand. The coexistence of mainland urban areas, transport hubs, residential zones, and tourist-specific regions creates heterogeneous travel patterns among the users across the city. These characteristics make Venice an interesting case study for exploring latent mobility behaviors and interactions among shared micromobility services under varying spatial and temporal conditions.

The user-based clustering methodology advances the micro mobility literature by introducing behavioral heterogeneity. Therefore, instead of just modeling trips, this study focuses on how heterogeneous mobility users can be identified from the high-dimensional categorical trip behavior while considering the uncertainty in the assignment of the cluster. Users are used as the fundamental unit of analysis and explores latent cluster patterns. In addition, it supports service providers and policy makers in identifying the demand and patterns of their customers, allowing them to optimize existing services and design stations. A repetitive pattern can be observed based on criteria such as the most-traveled zones or the choice of mode of transportation \parencite{veveEstimationSharedMobility2020}.

\section{Related Work}
The related literature analyzing the use of shared micromobility services has focused primarily on the demand structure at the trip-level, thus substantially improving the understanding of these transport services as a concept that supports the objectives of public health and urban livability \parencite{fishmanBikeShareSynthesis2013}. Further work has explored trip demands based on various factors, such as seasonal variability \parencite{beanHowDoesWeather2021,liSEASONALANALYSISFACTORS2017}, infrastructure \parencite{felixBuildItGive2020a}, and spatio-temporal \parencite{talavera-garciaExaminingSpatiotemporalMobility2021}. The existing work has highlighted trip flow patterns and provided important insights from a trip statistics perspective. However, it leaves a gap to study those who make these recurrent trips and exhibit behavioral heterogeneity. 

Shared micromobility users have recently been studied as the analysis unit and reveal distinct profiles among “super” users and occasional users \parencite{wintersWhoAreSuperusers2019} based on their demographic and income profiles. Studies in Lisbon, Madrid, and Cluj-Napoca identify distinct behavioral user clusters, including enthusiast, committed, and flexible young-adult users \parencite{adoreanClusteringUsersNonusers2026}. Survey-based studies have categorized users according to their trip characteristics, seasonal choice, vehicle modes, frequency of usage, and behavioral preferences \parencite{mohiuddinExaminingMarketSegmentation2024, pobudzeiUserSegmentationBased2024}. 

Hierarchical clustering and K-means are widely used in mobility segmentation studies, where users are represented by aggregated total trip frequency \parencite{pobudzeiUserSegmentationBased2024}, providing greater interpretability and computational feasibility. A study conducted in Kunming, China \parencite{liuInvestigatingUserPreferences2025} uses Hierarchical Density-Based Spatial Clustering of Noise Applications (HDBSCAN) to identify frequent travel user locations using shared cycling order data. Users were categorized into dockless bike-sharing (DBS)-dominant, balanced, and electric bike-sharing (EBS)-dominant clusters based on repeated usage patterns using predefined threshold-based user categories. It enables the identification of broad behavioral user groups relevant to transport planning and policy design \parencite{anableComplacentCarAddicts2005}. However, aggregating recurrent trips into user-level summaries reduces the specification of heterogeneous mobility patterns across space, time, seasonality, and vehicle choice and collapses the variability across individual trips. 

Recent studies have increasingly adopted latent and probabilistic segmentation in shared mobility systems to represent behavioral heterogeneity. The latent cluster analysis framework has been used to identify unobserved mobility profiles associated with travel preferences, modal usage, multinomial integration, and adoption attitude \parencite{adoreanClusteringUsersNonusers2026,gerzinicDriversBarriersIntegrating2025}. These approaches improve the representation of an overlapping behavioral structure by probabilistically assigning users to latent groups. However, many existing studies rely on survey-derived attributes \parencite{fuSharedMicromobilityMultimodal2025} or aggregated user-level summaries, which limit the representation of recurrent trip-level behavior.  

This study extends the existing segmentation of shared micro mobility services by developing a Bayesian mixture of product-multinomial models to identify latent user profiles from repeated categorical trip behavior directly from observed shared mobility records. The proposed framework integrates the origin-destination structure, vehicle choice, duration of trip, and seasonality to capture the variation in mobility behavior for bikes and e-bikes within Venice.

\section{Data Description}
Detailed trip-level data for regular bikes and e-bikes are used to examine user behavior clusters in the 2024 dataset within the Municipality of Venice. The overall data set includes travel details, including the spatial coordinates of the origin and destination, start and end times, user ID, pass group, ride time, and vehicle type. The data consists of 94,880 trips with regular bikes and 127,806 with e-bikes, providing a robust data set for analysis in both shared mobility services. 

The trip-level data set of shared micromobility services obtained from the Municipality of Venice is observed at the user level, with 16298 unique users throughout the year. It should be noted that users with only one observed trip were excluded, as these occasional users are mainly tourists, with trips towards and from specific areas, thus constituting a peculiar cluster, quite different from the others. After removing users with one trip, the number of individuals using micro-mobility services reduces to 11312.

\begin{figure}
    \centering
    \includegraphics[width=0.9\linewidth]{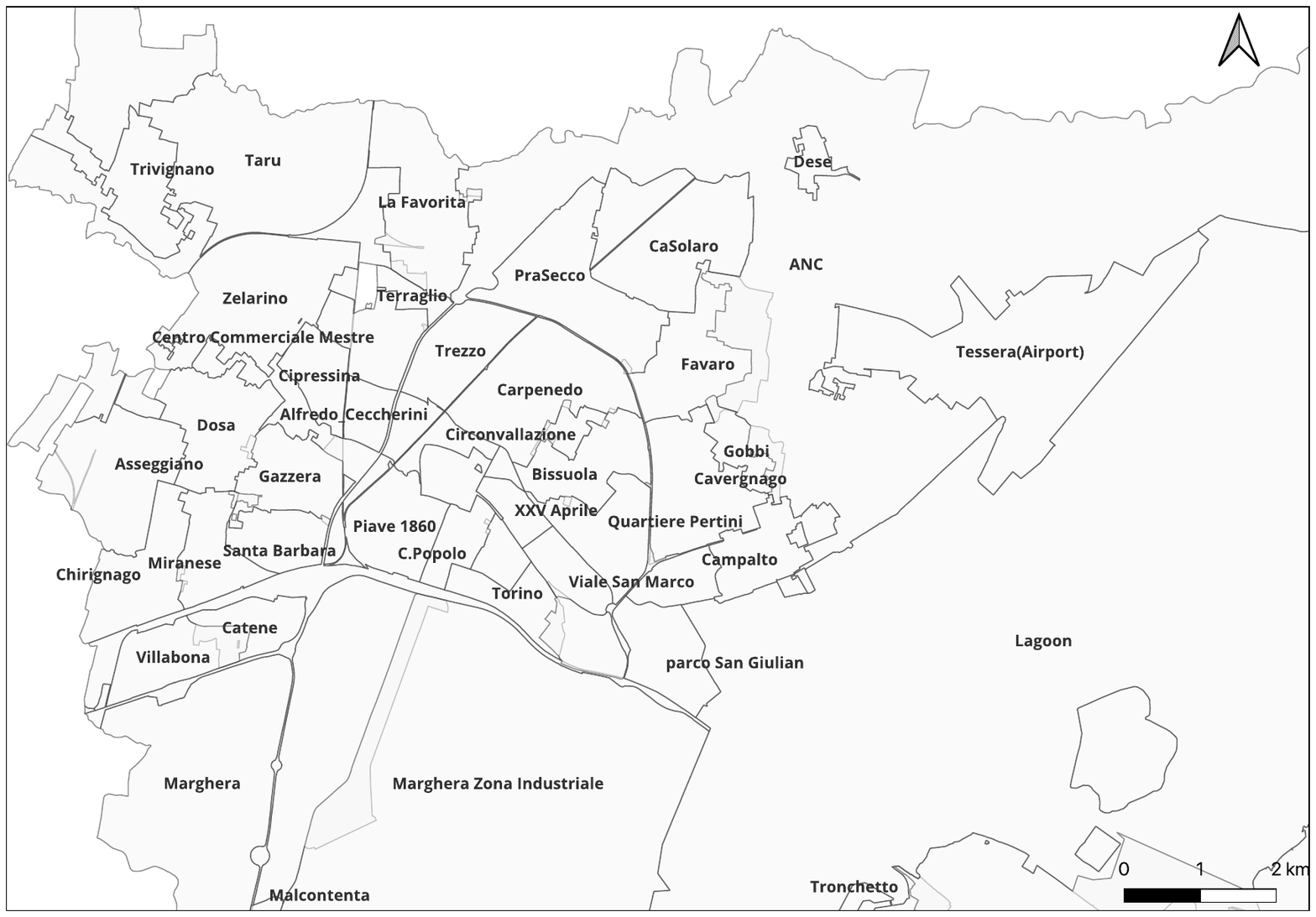}
    \caption{The mainland regions in the Municipality of Venice}
    \label{fig:Venicemap}
\end{figure}

The area of Venice is mainly divided into six administrative regions: Chiri\-gna\-go-Zelarino, Favaro Veneto, Lido-Pellestrina, Marghera, Mestre-Carpenedo, and Venice-Murano-Burano. Existing regions are further broken down into specific zones to analyze zone-specific usage patterns in QGIS using OpenStreetMap. The study area is then divided into 50 spatial zones according to the administrative and urban structure, as shown in the map in Figure \ref{fig:Venicemap}. Then, the travel origin and destination coordinates were spatially joined to these zones to aggregate the travels at the zone level. Figure ~\ref{fig:Venicemap} describes the formation of the zone layout in collaboration with the Venice Municipality. The zones included have distinct socio-spatial characteristics. For example, \emph{Piave 1860} and \emph{C.Popolo} are the main urban areas that include Mestre train station, while \emph{Torino} and \emph{Viale San Marco} are areas with some University buildings and municipal offices. Zones such as \emph{Marghera} are industrial, and \emph{Centro Commerciale Mestre} includes major commercial and healthcare facilities. The \emph{Lido} area, located in the southeastern part of the Municipality of Venice, includes Alberoni, Sandro Gallo, Elisabetta, San Nicolo, and Malamocco, which is predominantly characterized as a tourist area. 

The data set covers information on the distance covered for each trip. However, the distance covered has been considered unreliable. Therefore, a measure has been calculated using OpenStreetMap data to determine the most likely route and the related distance for the trip. The complete details of this method for calculating the distance are given in \parencite{padghamDodgrPackageNetwork2019}. Information on vehicle passes is available in six categories: coupon, monthly card \& daily pass, partner, pay-as-you-go (PAYG), premium pass, and times pass. The partner pass is only used for e-bikes, and the times pass as well, which is rarely used for bikes. The pass group category is combined with vehicle type, which is a bike and an e-bike. Table ~\ref{tab:vehicle_pass} shows the distribution of unique users and rides made across different vehicle types and pass categories. For example, 6,368 users made at least one bike trip using a PAYG bike pass throughout the year, representing 22,659 rides (24.6\% of all bike rides), while 8,269 users made at least one e-bike trip using a PAYG pass, accounting for 74,423 rides (60.4\% of all e-bike rides).





\begin{table}[!htbp]
\centering

\label{tab:vehicle_pass}

\begin{tabular}{lrrrr}
\hline
& \multicolumn{2}{c}{\textbf{Bike}}
& \multicolumn{2}{c}{\textbf{E-bike}}\\
\cline{2-3}\cline{4-5}

\textbf{Vehicle Pass}
& \textbf{Users}
& \textbf{Rides}
& \textbf{Users}
& \textbf{Rides}\\

& & \textbf{n (\%)}
& & \textbf{n (\%)}\\
\hline

Coupon                     & 73    & 78 (0.1)          & 111   & 119 (0.1) \\
Monthly Card \& Daily Pass & 999   & 64,287 (69.8)     & 143   & 501 (0.4) \\
Partner                    & --    & --                & 52    & 471 (0.4) \\
PAYG                       & 6,368 & 22,659 (24.6)     & 8,269 & 74,423 (60.4) \\
Premium Pass               & 780   & 5,077 (5.5)       & 1,141 & 22,494 (18.3) \\
Times Pass                 & 2     & 3 (0.0)           & 2,063 & 25,233 (20.5) \\

\hline
\end{tabular}
\caption{Distribution of unique users and rides across vehicle pass}
\end{table}

The data set also contains details of the ride time for each trip. We categorize the numerical data into five groups based on the distribution: less than 3 minutes, 3 to 6 minutes, 6 to 9 minutes, 9 to 12 minutes, and more than 12 minutes. Table 2 presents the distribution of users in each time category. It shows that among 11312 users, 27.9\% trips were made in 3-6 minutes, followed by 6-9 minutes, and more than 12 minutes.

\begin{table}[!htbp]
\centering

\label{tab:ride}

\begin{tabular}{lrr}
\hline
\textbf{Ride time} &
\textbf{Users} &
\textbf{Rides, n (\%)} \\
\hline

Less than 3 minutes  & 5,177 & 31,498 (14.6) \\
3--6 minutes         & 6,946 & 60,035 (27.9) \\
6--9 minutes         & 6,917 & 47,736 (22.2) \\
9--12 minutes        & 5,717 & 28,610 (13.3) \\
More than 12 minutes & 7,803 & 47,466 (22.0) \\

\hline
\end{tabular}
\caption{Distribution of unique users and rides across ride duration categories}
\end{table}

Using the trip's start and end dates, we categorized months into seasons to understand users' yearly preferences for sharing micromobility services. Table ~\ref{tab:season_users} includes the number of users who made trips during the four seasons. The highest number of users traveled in the summer with 33\% of rides, followed by autumn (27.5\%), spring (22.6\%), and winter (16.9\%). 

\begin{table}[!htbp]
\centering

\label{tab:season_users}

\begin{tabular}{lrr}
\hline
\textbf{Season} &
\textbf{Users} &
\textbf{Rides, n (\%)} \\
\hline

Autumn  & 5,286 & 59,205 (27.5) \\
Spring  & 4,357 & 48,620 (22.6) \\
Summer  & 7,059 & 71,161 (33.0) \\
Winter  & 3,679 & 36,359 (16.9) \\

\hline
\end{tabular}
\caption{Distribution of unique users and rides across seasons}
\end{table}
Unfortunately, the service provider does not collect user-specific information, such as age and gender, and thus it is unavailable.

\section{Methodology}

Each individual's record is considered instead of collapsing data based on its dominant features. Let \(N\) denote the number of users, for each user \(i = 1, \ldots, N\), who made various trips that are described by multiple categorical attributes, such as origin, destination, vehicle pass, season, and travel duration. The vehicle pass category is the combination of vehicle type and pass group variables.

Each user is represented as a set of categorical observations in a set of \(M\) attributes. For each attribute \(m\), let \(y_i^{(m)} \in \mathbb{N}^{C_m}\) denote the vector of counts over \(C_m\) categories. To capture heterogeneity in user travel patterns, it is assumed that each user belongs to one of \(K\) latent clusters, each representing distinct mobility profiles across trip attributes, consistent with a latent-class and mixture-model structure for categorical observations \parencite{mclachlanFiniteMixtureModels2000}. It can be represented as follows:

\begin{equation}
z_i \sim \text{Categorical}(\pi), \quad i = 1, \ldots, N
\end{equation}

where \(\pi = (\pi_1, \ldots, \pi_K)\) are the mixture proportions. The observed attributes are conditional on cluster assignment \(z_i = k\), which is the latent cluster assignment for the user, and are modeled as independent multinomial distributions.

\begin{equation}
y_i^{(m)} \mid z_i = k \sim \text{Multinomial}\left(n_i^{(m)}, \theta_k^{(m)}\right), \quad m = 1, \ldots, M
\end{equation}

where

\begin{equation}
n_i^{(m)} = \sum_{c=1}^{C_m} y_{ic}^{(m)}
\end{equation}

is the total number of observations for the attribute \(m\) for user \(i\), where \(y_{ic}^{(m)}\) represents the number of trips made by a user belonging to the category of attribute and

\begin{equation}
\theta_k^{(m)} =
\left(
\theta_{k1}^{(m)},
\ldots,
\theta_{kC_m}^{(m)}
\right)
\end{equation}

is a cluster-specific probability vector over categories of attribute \(m\) for cluster \(k\). Therefore, the explicit multinomial likelihood for attribute \(m\) is:

\begin{equation}
p\left(
y_i^{(m)}
\mid
z_i = k,
\theta_k^{(m)}
\right)
=
\frac{
\left(n_i^{(m)}\right)!
}{
\prod_{c=1}^{C_m} y_{ic}^{(m)}!
}
\prod_{c=1}^{C_m}
\left(
\theta_{kc}^{(m)}
\right)^{y_{ic}^{(m)}}
\end{equation}

assuming conditional independence across attributes given the cluster \parencite{bleiLatent_Dirichlet_Allocation2003,mclachlanFiniteMixtureModels2000}, the likelihood contribution of the user \(i\) is written as the product of multinomial likelihoods:

\begin{equation}
p(y_i \mid z_i = k)
=
\prod_{m=1}^{M}
p\left(
y_i^{(m)}
\mid
\theta_k^{(m)}
\right)
\end{equation}

It defines a product multinomial likelihood. The multinomial distribution is a probabilistic model for user-level counts, and the product-multinomial structure allows trip attributes to retain separate categories while sharing a latent cluster allocation.

The conditional independence assumption implies that, once the user’s latent cluster is identified, the attribute-specific count vectors are modeled independently. It is adopted in a latent-class and mixture-of-multinomials framework to obtain a tractable representation of high-dimensional categorical behavior.

A Dirichlet prior is placed in the model parameters as:

\begin{equation}
\pi \sim \text{Dirichlet}(\alpha_0),
\qquad
\theta_k^{(m)} \sim \text{Dirichlet}(\beta_0)
\end{equation}

where \(\alpha_0\) and \(\beta_0\) are hyperparameters controlling sparsity and smoothing. Values smaller than \(1\) favor sparse cluster formation and concentrated category probabilities, leading to more distinct latent clusters, while values greater than \(1\) favor a more uniform distribution across categories. We specify \(\alpha_0\) and \(\beta_0\) as \(0.5\) based on checking them at different values (sensitivity analysis). 

Upon combining the conditional independence between the marginal components of the likelihood and the prior distribution of the model, the marginal likelihood for the model can be written as:

\begin{equation}
p(y,z,\pi,\theta)
=
\left(
\prod_{i=1}^{N}
\prod_{m=1}^{M}
p\left(
y_i^{(m)}
\mid
z_i,\theta
\right)
\right)
\left(
\prod_{i=1}^{N}
p(z_i \mid \pi)
\right)
p(\pi)p(\theta)
\end{equation}

In other words, it is the full joint distribution of the observed user-level count data \(y\), latent cluster assignments \(z\), the mixture proportions \(\pi\), and the cluster-specific attribute parameters \(\theta\).

The proposed framework belongs to the family of finite mixture models for multivariate categorical data. It is related to latent class models and mixture-of-multinomials \parencite{bleiLatent_Dirichlet_Allocation2003} as well as the approach widely used in text mining, genetics, and behavioral profiling.

\begin{figure}[H]
    \centering
    \includegraphics[width=1.0\linewidth]{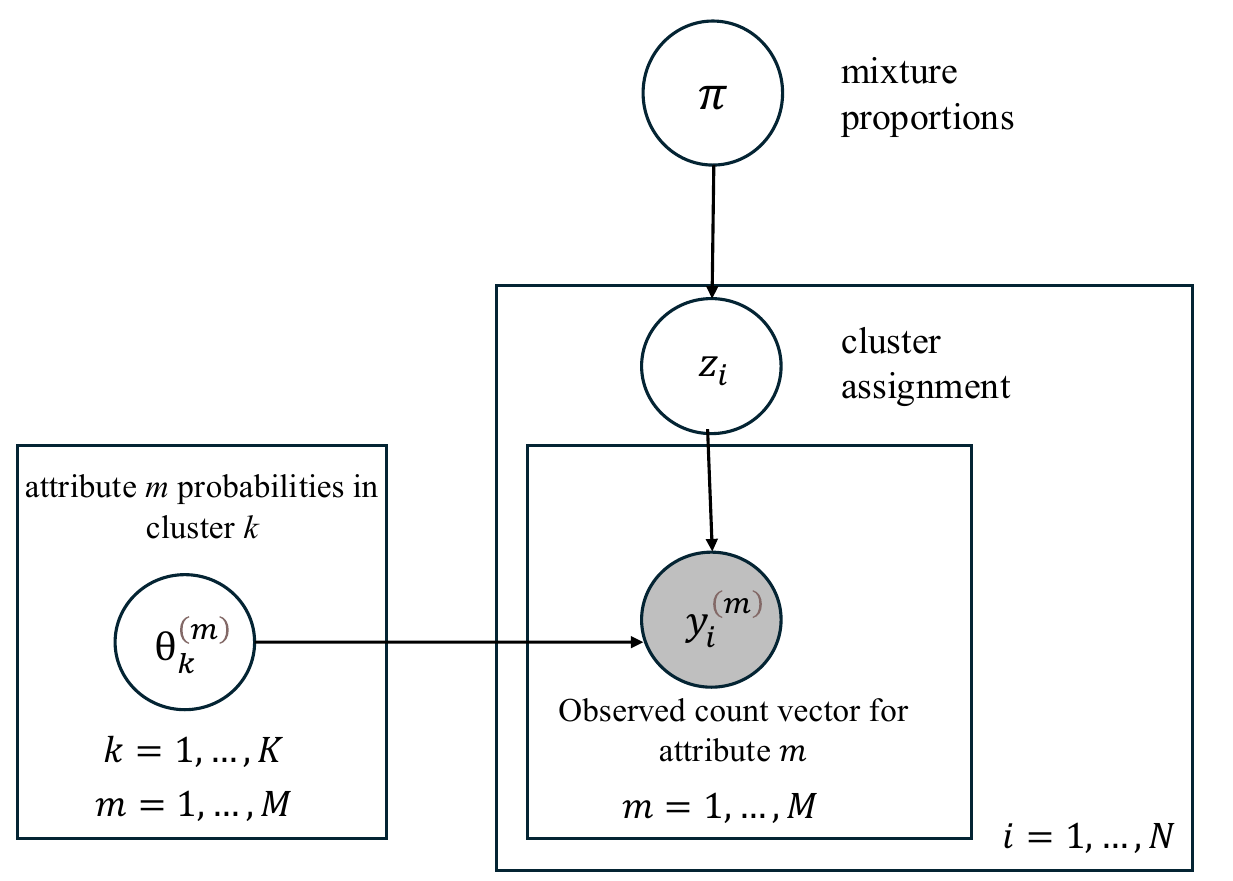}
    \caption{Directed Acyclic Graph (DAG) representation of the product-multinomial mixture model for user-level mobility behavior. The circles represent random variables, arrows denote probabilistic dependence, and shaded nodes correspond to observed data. Plates indicate replication across users, attributes, and clusters. The latent cluster assignment \(z_i\) controls the generation of attribute-specific count vectors  \(y_i^{(m)}\), with cluster-specific multinomial parameters \(\theta_k^{(m)}\).}
    \label{fig:plate_model}
\end{figure}

Figure~\ref{fig:plate_model} provides a graphical representation of the model. The proportions of the mixture \(\pi\) determine the number of users assigned to each cluster \(z_i\), deciding on the generation of user-level attribute counts \(y_i^{(m)}\) through a cluster-specific multinomial parameter \(\theta_k^{(m)}\). Plates indicate replication between users \(i = 1, \ldots, N\), attributes \(m = 1, \ldots, M\), and clusters \(k = 1, \ldots, K\).

Since Bayesian estimation requires computing the posterior distribution,\\ 
\(p(z,\pi,\theta \mid y)\), which is proportional to the joint distribution, the exact inference is intractable because it depends on the marginal likelihood, which involves integrating over the parameters and summing over all possible latent cluster assignments:

\begin{equation}
p(y \mid \alpha_0,\beta_0)
=
\int
\sum_{z}
p(y \mid z,\theta)\,
p(z \mid \pi)\,
p(\theta)\,
p(\pi)
\, d\theta \, d\pi
\end{equation}

Clearly, the posterior distribution cannot be computed analytically. Moreover, MCMC methods scale poorly in this case, and considering the large number of users and the relatively high dimension of categorical responses, we need to adopt a variational inference approach \parencite{bleiVariationalInferenceReview2017,jordanIntroductionVariationalMethods1998}. This allows for approximate calculation of the true posterior with a tractable distribution \(q=(\pi,\theta,z)\). To implement this approach, a classical mean-field approximation is chosen \parencite{bleiVariationalInferenceReview2017,wainwrightGraphicalModelsExponential2008}, assuming that the variational distribution factorizes as:

\begin{equation}
q(\pi,\theta,z)
=
q(\pi)
\prod_{k=1}^{K}
\prod_{m=1}^{M}
q\left(\theta_k^{(m)}\right)
\prod_{i=1}^{N}
q(z_i)
\end{equation}

where \(q(z_i)\) is a categorical distribution with parameters \(\phi_i\), and \(q(\pi)\) and \(q\left(\theta_k^{(m)}\right)\) follow Dirichlet distributions with variational parameters \(\alpha\) and \(\beta_k^{(m)}\), respectively.

The variational parameters are estimated by maximizing the Evidence Lower Bound (ELBO), defined as:

\begin{equation}
\mathcal{L}(q)
=
\mathbb{E}_q
\left[
\log p(y,z,\pi,\theta)
\right]
-
\mathbb{E}_q
\left[
\log q(\pi,\theta,z)
\right]
\end{equation}

which is a functional of the variational distribution \(q\) and provides a lower bound on the marginal log-likelihood. Maximizing ELBO is equivalent to minimizing the Kullback-Leibler divergence \parencite{murphyMachineLearningProbabilistic2013a} between the variational distribution and the true posterior.

An algorithm for maximizing the ELBO is performed using coordinate ascent variational inference (CAVI), where the ELBO is maximized for one parameter at a time, holding the others constant, and iteratively updating the estimates until convergence is achieved \parencite{leeGibbsSamplerCoordinate2022}. In particular, the cluster assignment probabilities are updated according to

\begin{equation}
\phi_{ik}
\propto
\exp
\left(
\mathbb{E}_q[\log \pi_k]
+
\sum_{m=1}^{M}
\sum_{c=1}^{C_m}
y_{ic}^{(m)}
\mathbb{E}_q
\left[
\log \theta_{kc}^{(m)}
\right]
\right)
\end{equation}

where \(\phi_{ik}\) represents the posterior probability that the user \(i\) belongs to cluster \(k\), computed based on the expected log mixture proportions and weighted by the observed counts for each attribute. The variational parameters are updated using the expected sufficient statistics under the current variational distribution. The Dirichlet parameters for the mixture proportions and attribute distributions are updated using expected cluster memberships.

To know the appropriate number of clusters, the value of \(K\) is estimated in the range of \(K \in \{5, \ldots, 12\}\) and the optimal value is selected by comparing ELBO values, together with cluster interpretability, entropy, and size stability to avoid over-fragmentation into very small clusters. Cluster entropy assessed assignment uncertainty and the balance between cluster sizes to avoid small-cluster fragmentation.

The final cluster assignment for each user is obtained by assigning each \(i\) to the cluster with the highest posterior membership probability, \(\phi_{ik}\)

\begin{equation}
\hat{z}_i
=
\arg\max_k \phi_{ik}
\end{equation}

Directed mobility graphs are used to visualize the dominant origin-destination relationships within each cluster by the most frequent flows. Network layouts are constructed using the Fruchterman-Reingold force-directed algorithm \parencite{fruchterman1991graph}, which positions nodes according to connectivity patterns to represent the topology of the mobility network. This algorithm iteratively locates connected nodes close together and simultaneously separates all nodes through attractive and repulsive forces, producing a network that reflects the underlying connectivity structure. Zones with a stronger connection are clustered together, while zones with fewer flows are located apart to improve visual interpretation. 

The study uses a probabilistic latent class analysis (LCA) framework, with novelty represented by recurrent trip-level count distributions within a variational product-multinomial framework applied to shared mobility behavior.

\section{Results}

We obtain eight distinct user clusters based on five trip-level attributes: origin, destination, vehicle pass, season, and travel duration. The origin and destination attributes comprised 50 categories, vehicle pass has 11 (we combine vehicle types, such as bike and ebike, and 6 types of pass groups, such as coupon, monthly card \& daily, PAYG, premium, times, and partner), season has been classified into 4 categories, and time duration has 5. 

\begin{table}[H]
\centering
\begin{tabular}{c c c c c c c c c}
\hline
\textbf{Clusters} & \textbf{1} & \textbf{2} & \textbf{3} & \textbf{4} & \textbf{5} & \textbf{6} & \textbf{7} & \textbf{8} \\
\hline
\textbf{Number of Users} 
& 1079 
& 1315 
& 4543 
& 424 
& 1837 
& 723 
& 906 
& 485 \\
\hline
\end{tabular}
\caption{Distribution of users in each cluster}
\label{tab:cluster_distribution}
\end{table}

They represent the heterogeneous mobility patterns in the Venice shared mobility system. Table ~\ref{tab:cluster_distribution} shows that the cluster sizes vary substantially, from the largest (Cluster 3) comprising 4,543 users to the smallest (Cluster 4) comprising 424 users. However, each cluster has sufficient users to be interpreted meaningfully. Apart from size, the cluster differs in its dominant vehicle type, seasonality, and trip duration, suggesting that multiple factors structure mobility behavior. 

In order to verify the ability of the proposed model to detect an underlying cluster of users, a simulation study has been implemented as a test. The details and results of simulation are presented in Appendix \ref{sec:simulation}, in practice more than 90\% of users have been correctly classified in the simulated data.

\subsection{Behavioral profiles across clusters}

\begin{figure}[H]
    \centering
    \includegraphics[width=1.0\linewidth]{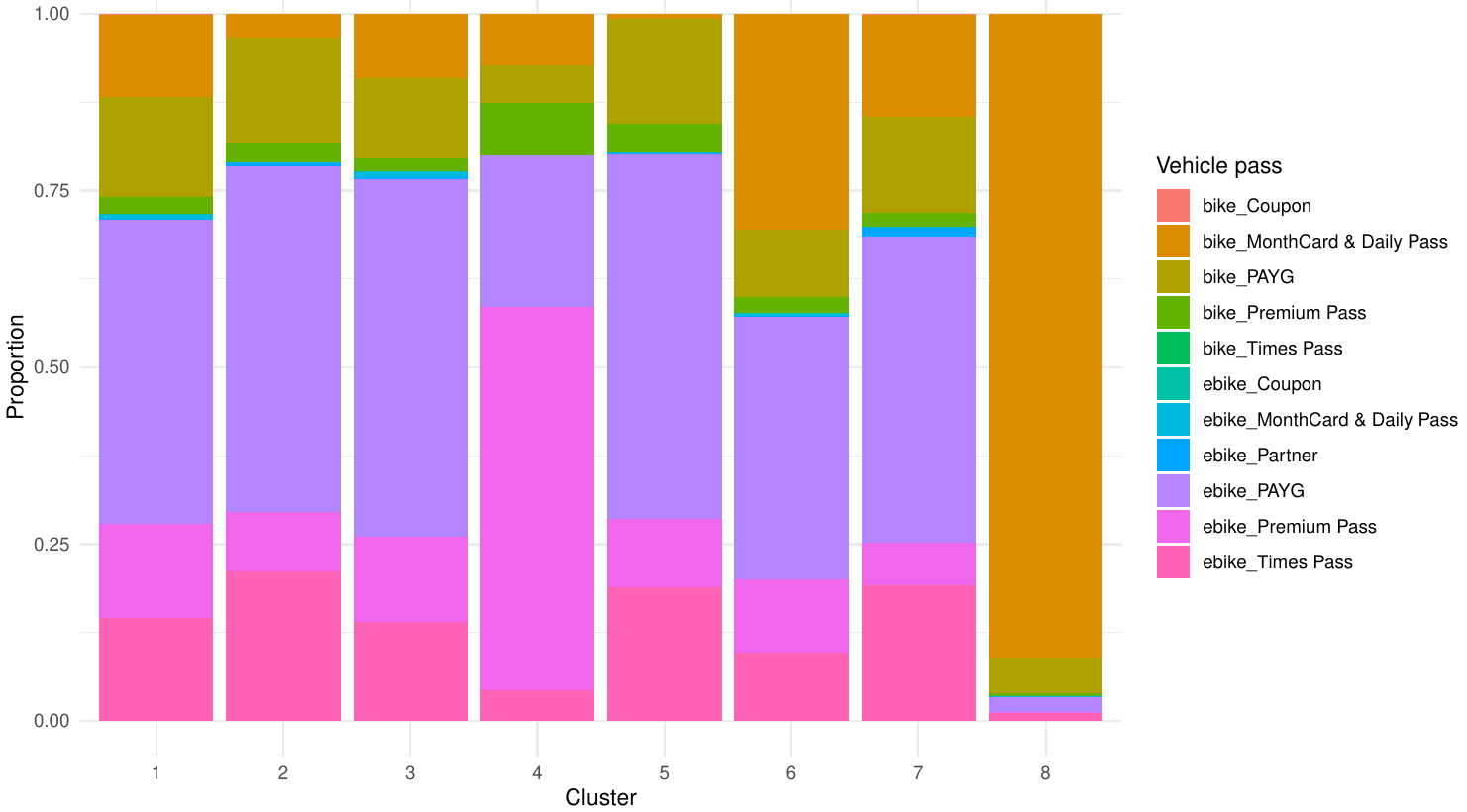}
    \caption{Vehicle pass distribution by cluster}
    \label{fig:vehicle_pass}
\end{figure}

Figures 2-4 present the distributions of vehicle pass, duration of trip, and seasonality in the groups, highlighting the variation within each group. Figure ~\ref{fig:vehicle_pass} shows the distribution of vehicle type by pass group across the cluster, reflecting the heterogeneity in subscription and vehicle mode among users. Most clusters are dominated by e-bike pass categories, particularly e-bike PAYG and time pass users. Clusters 2, 4, and 5 account for more than 75\% of users who opt for electric bikes, with PAYG being the most popular pass and the Premium pass specifically for Cluster 4. In contrast, Cluster 8 almost covers regular bike users (more than 90\%) with monthly \& daily passes, followed by Cluster 6, which also includes PAYG users, suggesting that vehicle type and pass type are key dimensions among users. 

\begin{figure}[H]
    \centering
    \includegraphics[width=1.0\linewidth]{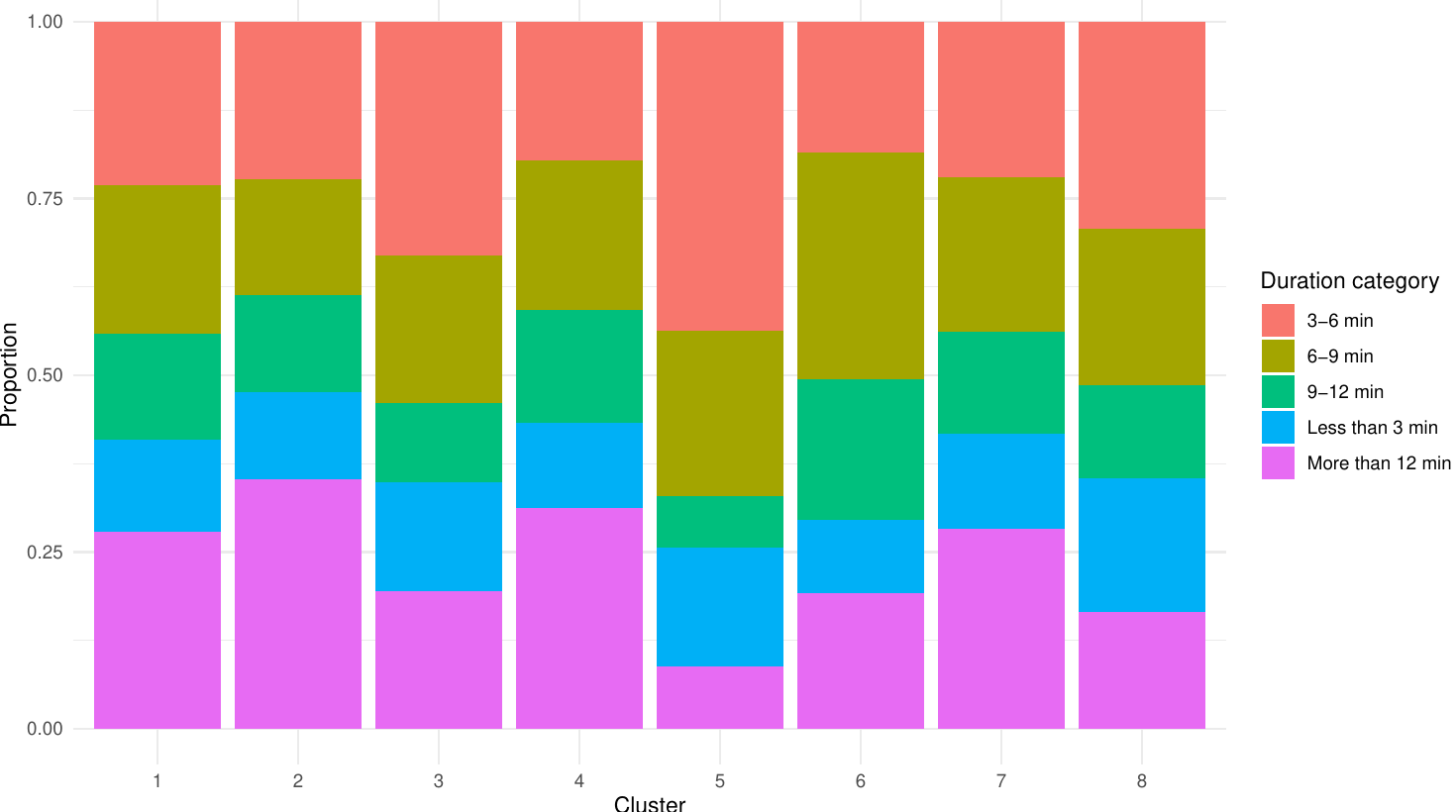}
    \caption{Trip-duration distribution by cluster}
    \label{fig:duration}
\end{figure}

The duration of the trip taken also reveals differences in the formed clusters. Figure ~\ref{fig:duration} shows that Clusters 3 and 5 are characterized by users who mostly take trips of shorter duration (3-6 minutes), indicating a local or last-mile usage pattern. In contrast, Clusters 1, 4, and 7 show a greater share of longer-duration trips (over 12 minutes), while Clusters 2 and 8 are more balanced in their duration profiles, indicating a mixed usage pattern. The factor of travel duration is strongly related to mobility purpose, which separates the commuting behavior of users into peripheral and sub-peripheral location coverage groups.

\begin{figure}[H]
    \centering
    \includegraphics[width=1.0\linewidth]{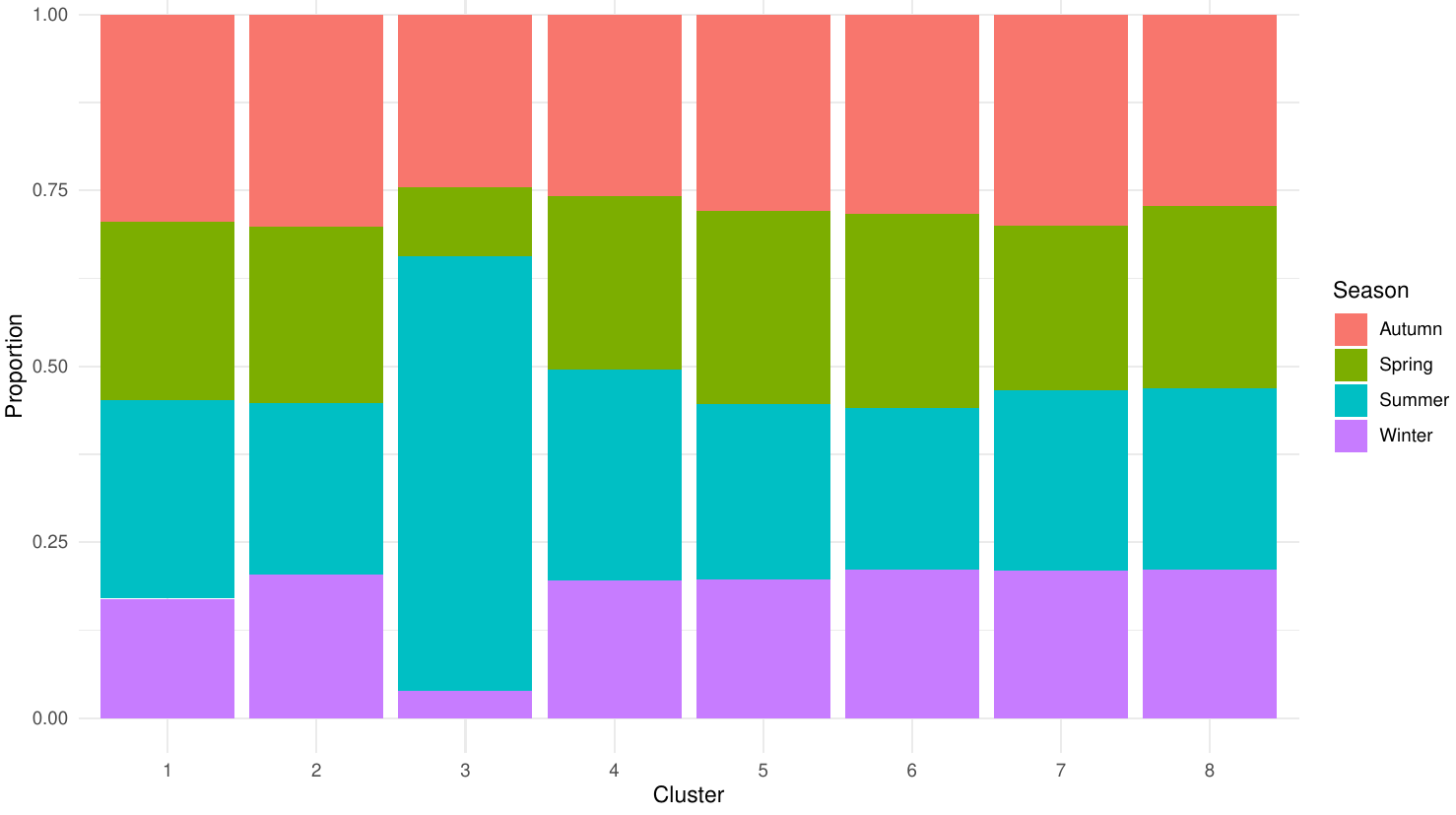}
    \caption{Seasonal distribution by cluster}
    \label{fig:season}
\end{figure}

Seasonality in determining user demand plays an integral role in understanding usage patterns across the year. Cluster 3 is strongly dominated by summer-season usage (Figure ~\ref{fig:season}), reflecting a tourist-based user cluster, particularly in the Lido area, such as Elisabetta, Alberoni, San Nicolo, and others. Other clusters are more balanced throughout the year, though there is a slight reduction during the winter season.

\subsection{Spatial structure of users’ clusters}

Spatial patterns represent the strongest differentiation across the cluster. The Fruchterman–Reingold force-directed approach layouts the cluster networks shown in Figure 6. In this representation, node positions are determined by the connectivity structure of the network. The spatial position of the nodes should not be interpreted as actual geographic distances or locations within Venice. They highlights the dominant origin–destination relationships and connectivity patterns within each user cluster. Figure ~\ref{fig:cluster_networks_2} shows the origin-destination flow networks for each cluster. Several distinct types of typologies can be inferred from the results obtained:

\begin{itemize}
\item Localized clusters: Cluster 1 mainly comprises users with trips around Marghera, with predominantly intra-zone flows, indicating a more local, work-related mobility cluster within the city’s industrial and employment areas. This cluster includes 1078 users, of whom 34\% travel within the Marghera region.
\item Centralized cluster: Clusters 2, 5, and 8 show the dominance of central areas such as Carpenedo, Piave 1860, Piazza Ferretto, and C.Popolo, indicating a corridor-based mobility pattern. These areas are the most represented in the user clustering, as they also include the Mestre railway station, which acts as a primary commuting hub connecting Venice with other urban and regional areas. In addition, it shows that the central area of the city is the main driving force of demand. It also includes mobility activities in the commercial part of the city, which includes a shopping complex and healthcare facilities.
\item Tourist-based cluster: Cluster 3 is strongly concentrated in the Lido zones (including L. Sandro Gallo and L. Elisabetta). It includes 4545 users, with 39\% of trips within Sandro Gallo. Although users who made only 1 trip were excluded from the analysis, which were mostly within Lido, even then, a high number of users fell into this cluster, indicating a higher usage of these services within the tourist-oriented area.
\item Inter-zone cluster: Clusters 4, 6, and 7 link multiple zones, including central, peripheral, and semi-peripheral areas (e.g., Viale San Marco, Gobbi, Cipresina, Santa Barbara, and others). These clusters reflect network-based mobility, which connects different parts of the city, including workplaces such as the university and municipal offices.
\end{itemize}

\begin{figure}[H]
\centering
\begin{tabular}{|c|c|}
\hline
\textbf{Cluster 1} & \textbf{Cluster 2} \\
\includegraphics[width=0.50\textwidth]{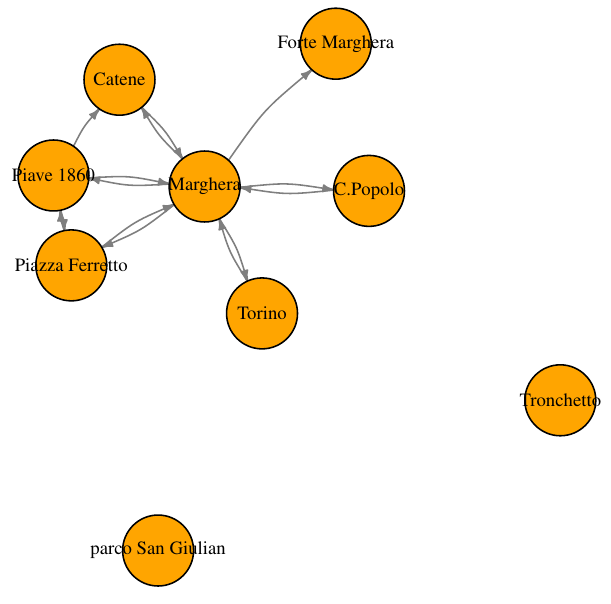} &
\includegraphics[width=0.50\textwidth]{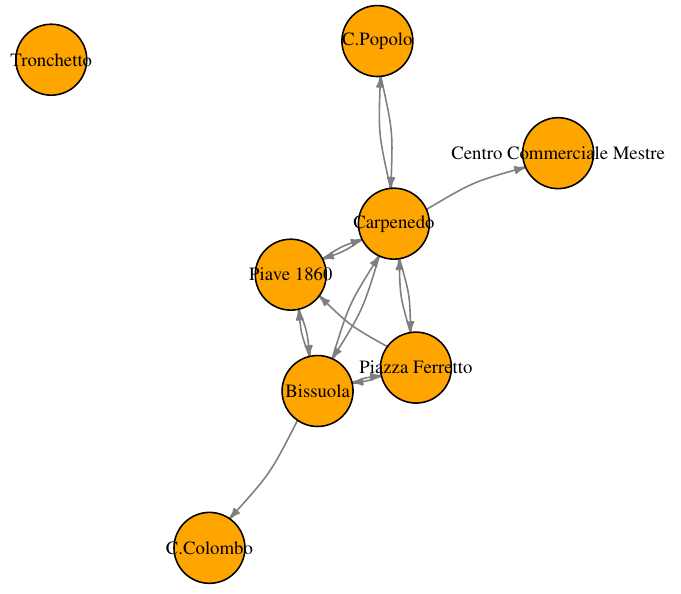} \\
\hline

\textbf{Cluster 3} & \textbf{Cluster 4} \\
\includegraphics[width=0.50\textwidth]{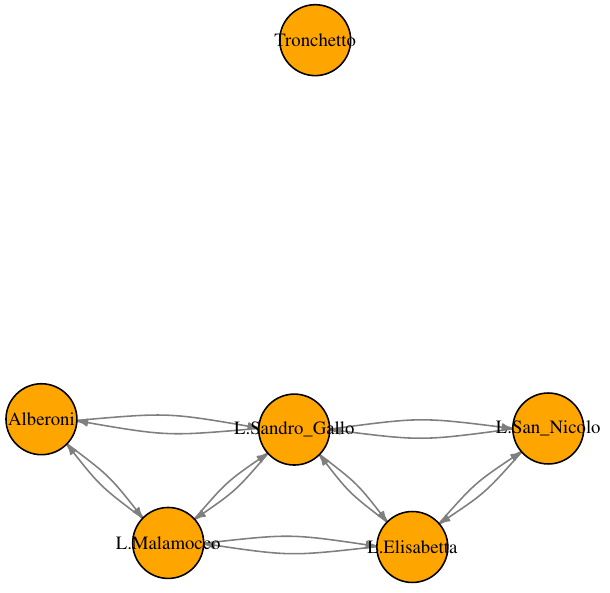} &
\includegraphics[width=0.50\textwidth]{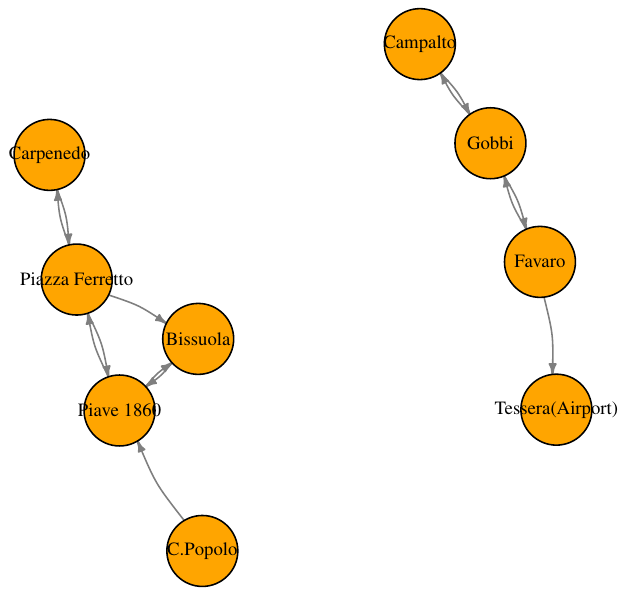}\\
\hline
\end{tabular}

\end{figure}

\begin{figure}[H]
\centering
\begin{tabular}{|c|c|}
\hline
\textbf{Cluster 5} & \textbf{Cluster 6} \\
\includegraphics[width=0.50\textwidth]{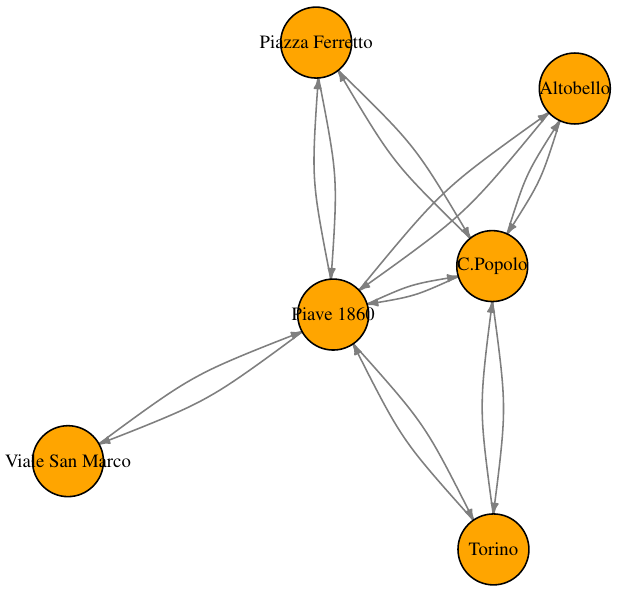} &
\includegraphics[width=0.50\textwidth]{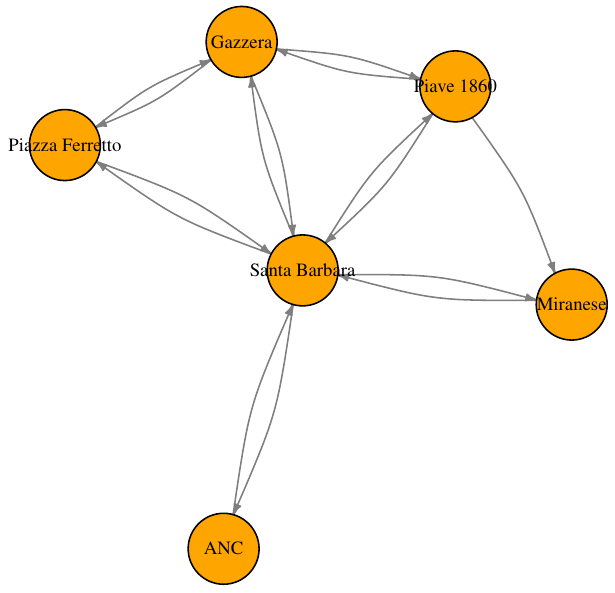} \\
\hline

\textbf{Cluster 7} & \textbf{Cluster 8} \\
\includegraphics[width=0.50\textwidth]{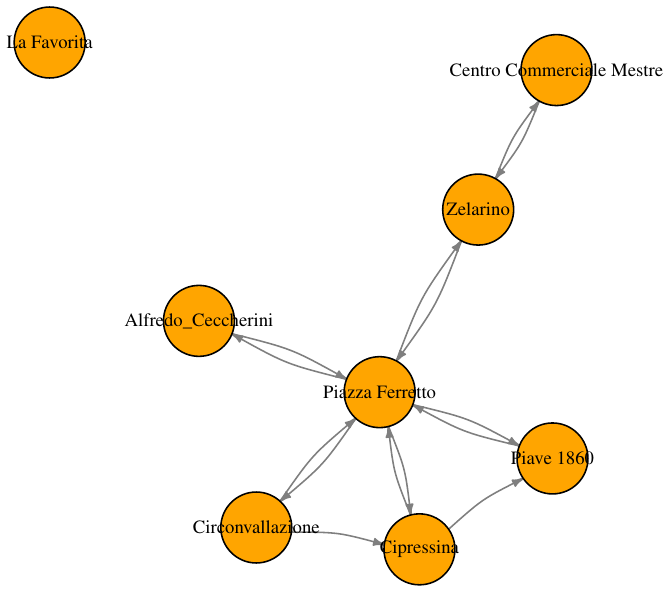} &
\includegraphics[width=0.50\textwidth]{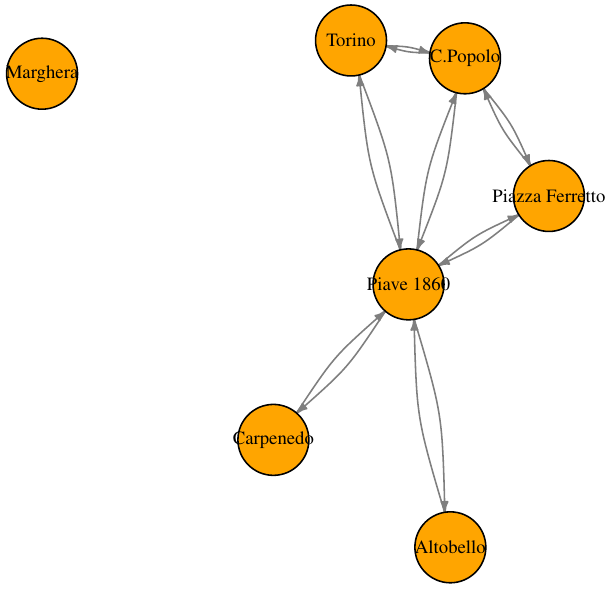}\\
\hline
\end{tabular}

\caption{Spatial network structures for latent mobility clusters}
\label{fig:cluster_networks_2}
\end{figure}

\subsection{Sensitivity analysis}
To assess the robustness of the clustering results, a sensitivity analysis was conducted for both the number of clusters \((K)\) and the prior hyperparameters of the variational Bayes model.

\begin{figure}[H]
    \centering
    \includegraphics[width=1.0\linewidth]{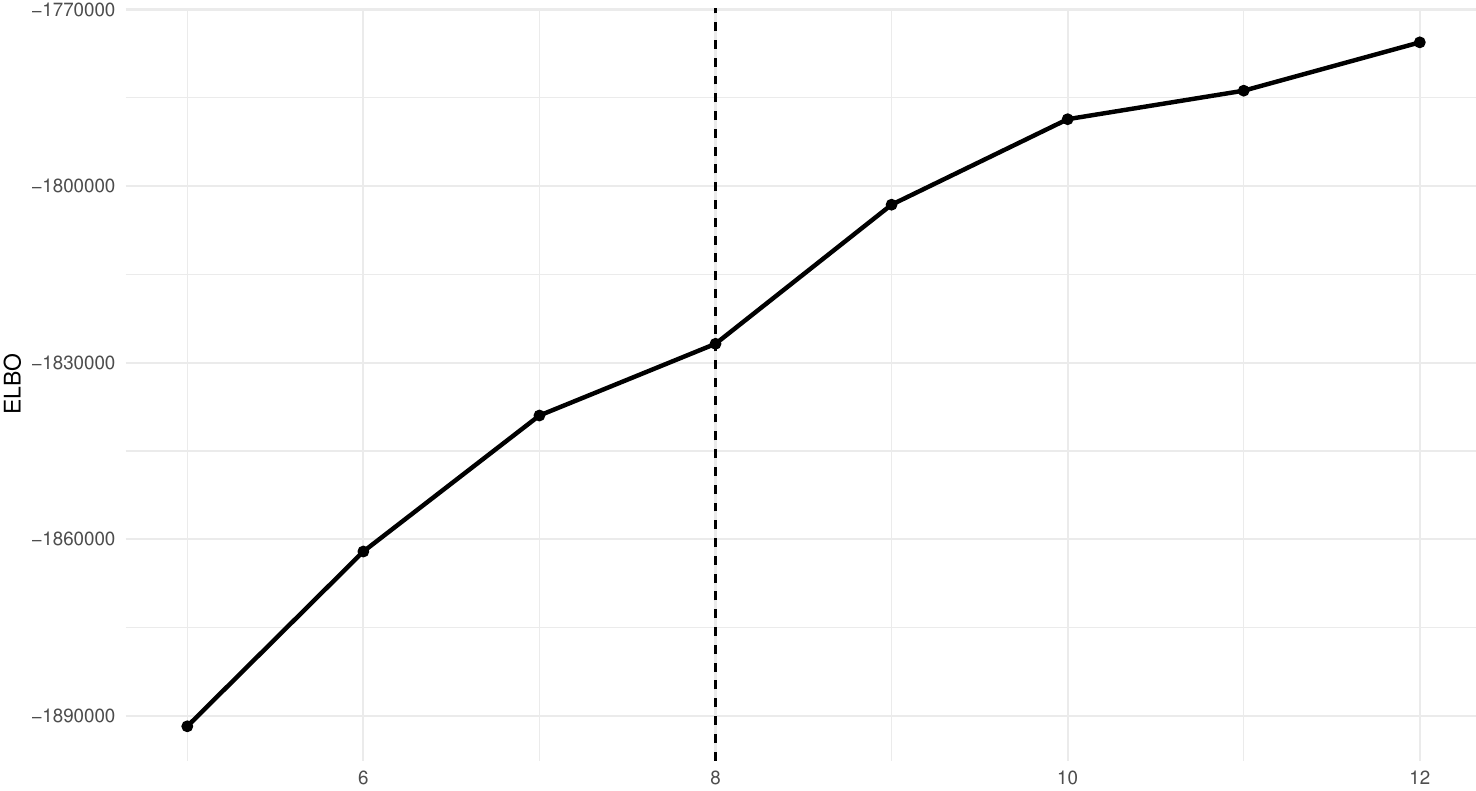}
    \caption{Variational Bayes model selection by ELBO}
    \label{fig:ELBO}
\end{figure}

The model is estimated across a range of latent cluster sizes 
\((K= 5:12)\), with multiple random initializations to maintain stability. Model selection is performed using the Evidence Lower Bound (ELBO). As shown in Figure ~\ref{fig:ELBO}, ELBO increases monotonically with \(K\), indicating an improved variational approximation. Although it continues to increase beyond\(K=8\), the incremental improvements become smaller, indicating diminishing returns from increasing complex models. Taking into account the trade-off between the complexity of the model, the interpretability of the clusters and the stability of the resulting solution, \(K=8\) was selected as the optimal number of clusters on the model selection curve.

\begin{figure}[H]
\centering
\includegraphics[width=1.0\textwidth]{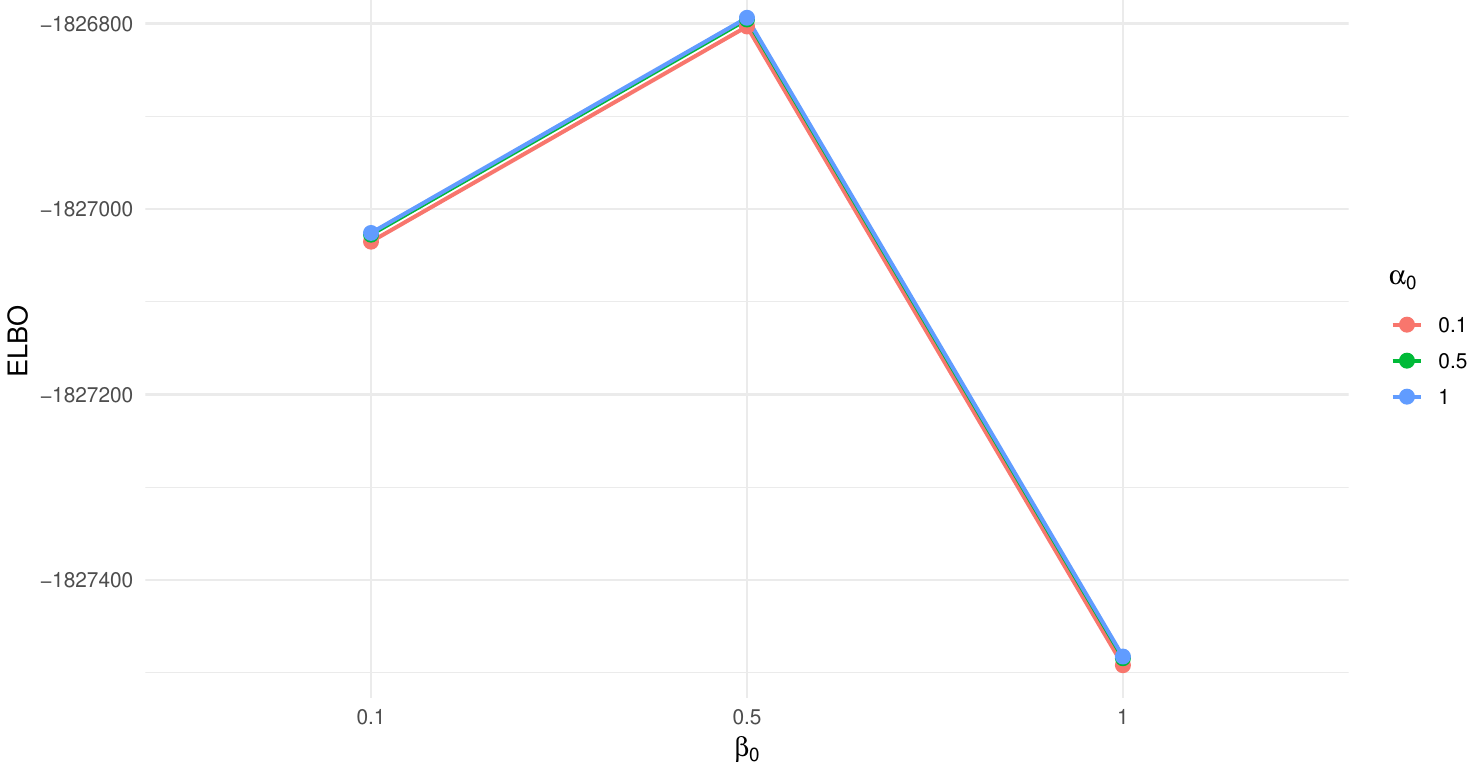}
\caption{Hyperparameter sensitivity analysis.}
\label{fig:elbo}
\end{figure}

The robustness of the clustering results was further evaluated by varying the Dirichlet prior parameters $\alpha_0$ and $\beta_0$ over the set $\{0.1, 0.5, 1\}$. Across all configurations, the resulting clustering structure remains the most stable. The number of clusters and cluster sizes showed minimal variation, with the smallest cluster ranging from 418 to 433 users and the largest cluster of 4538 to 4546 users. The ELBO values also showed minor differences of less than 0.05\% across parameter settings relative to the overall scale. The highest ELBO values were observed for $\beta_0 = 0.5$, suggesting that moderate smoothing of the category distributions provides the best fit (see  Figure~\ref{fig:elbo}). The cluster structure and cluster sizes remained stable in all specifications, indicating that the proposed variational product-multinomial framework is robust to moderate changes in prior assumptions.

\section{Discussion and Conclusion}

The study identifies eight distinct latent user clusters within the Venice shared micromobility system using a Bayesian mixture of multinomial product models estimated via variational approach. The resulting clusters show substantial heterogeneity between users in terms of vehicle pass, duration of travel, and season preference. Users traveling to different locations with homogeneous behavior are represented in the obtained clusters. The clusters are classified into localized, central, tourist-based, and inter-zone categories based on the spatial zones they cover within the studied area. The findings show that the shared demand for micromobility varies between users and follows a recurrent behavioral structure shaped by travel preferences, seasonal choices, and spatial demand patterns.  

Previous work has shown that micromobility users differ in travel frequency, preferred travel zones, vehicle choice, and seasonal usage patterns \parencite{gkartzonikasAssessmentTravelBehavior2026a,javaheriAnalyzingUsageTrends2025,pobudzeiUserSegmentationBased2024,roig-costaUnderstandingMultimodalMobility2026}. The localized structured pattern observed in Cluster 1 suggests a recurrent neighborhood pattern with dominant users traveling in Marghera. Similar behavior has been associated with routine commute and first and last-mile travel in shared mobility systems \parencite{beanHowDoesWeather2021,fanHowHaveTravelers2019}. The results for Cluster 3 show a strong preference for summer, characterized by users in the Lido region (a more visited tourist area). The results align with the tourism-oriented mobility behavior during peak periods \parencite{christoforouNeighborhoodCharacteristicsEncouraging2023}. Clusters 4, 5 and 6 show inter-zonal connections, linking peripheral, semi-peripheral, and central areas, connecting different parts of the municipality. Such patterns suggest that micromobility services are used for a larger network in the city. Although studies comparing bike-sharing and other modes show that the latter are associated with dominant use in peripheral areas \parencite{adoreanClusteringUsersNonusers2026,liuSharedMicromobilityMeets2025}, micromobility serves primarily as first-mile trips \parencite{bezaSpatiotemporalAnalysisShared2025}.

Clusters reflect significant variations in the type of pass combined with vehicle mode. The Pay-as-you-Go category (PAYG) dominated the most clusters, specifically combined with e-bikes, followed by monthly \& daily bike passes and time passes for e-bikes. The committed users who hold the monthly card are predominantly regular bike users, while casual users appear to be more likely to use electric bikes. Similar patterns reflecting daily commute behavior through regular bike-sharing membership have been observed in Washington, D.C.\parencite{mckenzieSpatiotemporalComparativeAnalysis2019}. At the same time, scooters do not support this standard commuting behavior and are used primarily for leisure or tourism. 

Unlike previous approaches to clustering users that rely on aggregating trip frequencies and using summary-based measures, this study proposes a framework that identifies recurrent individual trips and leverages multiple categorical user-level attributes. Traditional clustering methods, such as hierarchical clustering, latent class analysis, or K-means, produce latent user clusters that collapse user behavior to their dominant characteristics, such as average travel distance or total travel frequency \parencite{ferrariExtractingUrbanPatterns2011,gerzinicDriversBarriersIntegrating2025,roig-costaUnderstandingMultimodalMobility2026}. This study uses the Bayes product multinomial variational framework to probabilistically model the latent behavioral structure of recorded trips with different patterns observed in origin, destination, vehicle-pass, travel duration, and seasonality. It allows for the representation of mixed behavioral characteristics alongside latent mobility profiles. 

In a policy-oriented context, the research uses real-world data rather than survey-based preferences, thereby accounting for the practical use of shared micromobility services. The identification of tourist, localized, central, and inter-zonal clusters highlights the importance of allocating services based on seasonal and location preferences. For example, more bikes and electric bikes can be installed in the Lido region during the summer season, as demand is high then. In addition, the inter-zonal clusters indicate that shared micromobility services are connecting peripheral and central districts, supporting corridor-based urban connectivity. Since the PAYG pass group category has the highest usage among e-bike users and the monthly bike pass is also popular among bike users, introducing similar passes can increase adoption of these services. Studying trips within a probabilistic clustering framework would be a useful approach to user segmentation while preserving each user's repeated trip records. 

The study also has limitations, such as the analysis being restricted to bikes and e-bikes and not accounting for other modes such as e-scooters or public transit. Additionally, the product multinomial framework assumes conditional independence of attributes within clusters, thereby ignoring attribute correlations; however, this does not affect our cluster solutions. Future work could examine the socio-economic characteristics of bike users, including age, income, occupation, and education level. It would help to define the correlation between the type of pass and the other behaviors of different groups.

\printbibliography

@article{adoreanClusteringUsersNonusers2026,
  title = {Clustering Users and Non-Users of Shared and Private e-Bikes and e-Scooters across Three {{European}} Cities: {{Typologies}}, Spatial Patterns, and Policy Implications},
  shorttitle = {Clustering Users and Non-Users of Shared and Private e-Bikes and e-Scooters across Three {{European}} Cities},
  author = {Adorean, Emanuel-Cristian and Nofre, Jordi and Moura, Filipe},
  date = {2026-01},
  journaltitle = {Transport Policy},
  volume = {175},
  pages = {103875},
  issn = {0967070X},
  doi = {10.1016/j.tranpol.2025.103875},
  url = {https://linkinghub.elsevier.com/retrieve/pii/S0967070X25004184},
}

@article{anableComplacentCarAddicts2005,
  title = {‘{{Complacent Car Addicts}}’ or ‘{{Aspiring Environmentalists}}’? {{Identifying}} Travel Behaviour Segments Using Attitude Theory},
  shorttitle = {‘{{Complacent Car Addicts}}’ or ‘{{Aspiring Environmentalists}}’?},
  author = {Anable, Jillian},
  date = {2005-01},
  journaltitle = {Transport Policy},
  volume = {12},
  number = {1},
  pages = {65--78},
  issn = {0967070X},
  doi = {10.1016/j.tranpol.2004.11.004},
  url = {https://linkinghub.elsevier.com/retrieve/pii/S0967070X0400054X},
}

@article{beanHowDoesWeather2021,
  title = {How Does Weather Affect Bikeshare Use? {{A}} Comparative Analysis of Forty Cities across Climate Zones},
  shorttitle = {How Does Weather Affect Bikeshare Use?},
  author = {Bean, Richard and Pojani, Dorina and Corcoran, Jonathan},
  date = {2021-07},
  journaltitle = {Journal of Transport Geography},
  volume = {95},
  pages = {103155},
  issn = {09666923},
  doi = {10.1016/j.jtrangeo.2021.103155},
  url = {https://linkinghub.elsevier.com/retrieve/pii/S0966692321002088},
}

@article{bezaSpatiotemporalAnalysisShared2025,
  title = {A {{Spatiotemporal Analysis}} of {{Shared Micromobility Trips}} in {{First-}} and {{Last-Mile Public Transit Integration}}},
  author = {Beza, Abebe Dress and Demissie, Merkebe Getachew and Kattan, Lina},
  date = {2025-11},
  journaltitle = {Transportation Research Record: Journal of the Transportation Research Board},
  volume = {2679},
  number = {11},
  pages = {762--781},
  issn = {0361-1981, 2169-4052},
  doi = {10.1177/03611981251350652},
  url = {https://journals.sagepub.com/doi/10.1177/03611981251350652},
}

@article{bleiLatent_Dirichlet_Allocation2003,
  title = {Latent\_{{Dirichlet}}\_{{Allocation}}},
  author = {Blei, David M. and Ng, Andrew Y. and Jordan, Michael I.},
  date = {2003},
  journaltitle = {CrossRef Listing of Deleted DOIs},
  volume = {3},
  issn = {0003-6951},
  doi = {10.1162/jmlr.2003.3.4-5.993},
  url = {https://www.jmlr.org/papers/volume3/blei03a/blei03a.pdf?ref=http://githubhelp.com},
}

@article{bleiVariationalInferenceReview2017,
  title = {Variational {{Inference}}: {{A Review}} for {{Statisticians}}},
  shorttitle = {Variational {{Inference}}},
  author = {Blei, David M. and Kucukelbir, Alp and McAuliffe, Jon D.},
  date = {2017-04-03},
  journaltitle = {Journal of the American Statistical Association},
  volume = {112},
  number = {518},
  pages = {859--877},
  issn = {0162-1459, 1537-274X},
  doi = {10.1080/01621459.2017.1285773},
  url = {https://www.tandfonline.com/doi/full/10.1080/01621459.2017.1285773},
}

@article{christoforouNeighborhoodCharacteristicsEncouraging2023,
  title = {Neighborhood Characteristics Encouraging Micromobility: {{An}} Observational Study for Tourists and Local Users},
  shorttitle = {Neighborhood Characteristics Encouraging Micromobility},
  author = {Christoforou, Zoi and Psarrou Kalakoni, Anna Mariam and Farhi, Nadir},
  date = {2023-07},
  journaltitle = {Travel Behaviour and Society},
  volume = {32},
  pages = {100564},
  issn = {2214367X},
  doi = {10.1016/j.tbs.2023.02.002},
  url = {https://linkinghub.elsevier.com/retrieve/pii/S2214367X23000091},
}

@article{fanHowHaveTravelers2019,
  title = {How {{Have Travelers Changed Mode Choices}} for {{First}}/{{Last Mile Trips}} after the {{Introduction}} of {{Bicycle-Sharing Systems}}: {{An Empirical Study}} in {{Beijing}}, {{China}}},
  shorttitle = {How {{Have Travelers Changed Mode Choices}} for {{First}}/{{Last Mile Trips}} after the {{Introduction}} of {{Bicycle-Sharing Systems}}},
  author = {Fan, Aihua and Chen, Xumei and Wan, Tao},
  date = {2019-05-14},
  journaltitle = {Journal of Advanced Transportation},
  volume = {2019},
  pages = {1--16},
  issn = {0197-6729, 2042-3195},
  doi = {10.1155/2019/5426080},
  url = {https://www.hindawi.com/journals/jat/2019/5426080/},
}

@article{felixBuildItGive2020a,
  title = {Build It and Give ‘em Bikes, and They Will Come: {{The}} Effects of Cycling Infrastructure and Bike-Sharing System in {{Lisbon}}},
  shorttitle = {Build It and Give ‘em Bikes, and They Will Come},
  author = {Félix, Rosa and Cambra, Paulo and Moura, Filipe},
  date = {2020-06},
  journaltitle = {Case Studies on Transport Policy},
  volume = {8},
  number = {2},
  pages = {672--682},
  issn = {2213624X},
  doi = {10.1016/j.cstp.2020.03.002},
  url = {https://linkinghub.elsevier.com/retrieve/pii/S2213624X20300183},
}

@inproceedings{ferrariExtractingUrbanPatterns2011,
  title = {Extracting Urban Patterns from Location-Based Social Networks},
  booktitle = {Proceedings of the 3rd {{ACM SIGSPATIAL International Workshop}} on {{Location-Based Social Networks}}},
  author = {Ferrari, Laura and Rosi, Alberto and Mamei, Marco and Zambonelli, Franco},
  date = {2011-11},
  pages = {9--16},
  publisher = {ACM},
  location = {Chicago Illinois},
  doi = {10.1145/2063212.2063226},
  url = {https://dl.acm.org/doi/10.1145/2063212.2063226},
  }

@article{fishmanBikeShareSynthesis2013,
  title = {Bike {{Share}}: {{A Synthesis}} of the {{Literature}}},
  shorttitle = {Bike {{Share}}},
  author = {Fishman, Elliot and Washington, Simon and Haworth, Narelle},
  date = {2013-03},
  journaltitle = {Transport Reviews},
  volume = {33},
  number = {2},
  pages = {148--165},
  issn = {0144-1647, 1464-5327},
  doi = {10.1080/01441647.2013.775612},
  url = {http://www.tandfonline.com/doi/abs/10.1080/01441647.2013.775612},
}

@article{fuSharedMicromobilityMultimodal2025,
  title = {Shared Micromobility in Multimodal Travel: {{Evidence}} from Three {{European}} Cities},
  shorttitle = {Shared Micromobility in Multimodal Travel},
  author = {Fu, Xingxing and Van Lierop, Dea and Ettema, Dick},
  date = {2025-03},
  journaltitle = {Cities},
  volume = {158},
  pages = {105664},
  issn = {02642751},
  doi = {10.1016/j.cities.2024.105664},
  url = {https://linkinghub.elsevier.com/retrieve/pii/S0264275124008783},
}

@article{gerzinicDriversBarriersIntegrating2025,
  title = {Drivers and Barriers to Integrating Shared Micromobility with Public Transport {{A}} Latent Class Clustering Analysis of Adoption Attitudes in the {{Netherlands}}},
  author = {Geržinič, Nejc and Van Hagen, Mark and Al-Tamimi, Hussein and Van Oort, Niels and Duives, Dorine},
  date = {2025-12},
  journaltitle = {Journal of Cycling and Micromobility Research},
  volume = {6},
  pages = {100090},
  issn = {29501059},
  doi = {10.1016/j.jcmr.2025.100090},
  url = {https://linkinghub.elsevier.com/retrieve/pii/S2950105925000348},
  }

@article{gkartzonikasAssessmentTravelBehavior2026a,
  title = {Assessment of Travel Behavior Dynamics among Young Demographic for Different Trip Purposes of Shared Micro-Mobility Services},
  author = {Gkartzonikas, Christos and Dimitriou, Loukas},
  date = {2026-05},
  journaltitle = {Transport Policy},
  volume = {180},
  pages = {104042},
  issn = {0967070X},
  doi = {10.1016/j.tranpol.2026.104042},
  url = {https://linkinghub.elsevier.com/retrieve/pii/S0967070X26000521},
  }

@article{javaheriAnalyzingUsageTrends2025,
  title = {Analyzing {{Usage Trends}} of {{Shared Micromobility Among University Students}}},
  author = {Javaheri, Atusa and Pamidimukkala, Apurva and Kermanshachi, Sharareh and Rosenberger, Jay Michael and Hladik, Greg},
  date = {2025},
  journaltitle = {Transportation Research Procedia},
  volume = {91},
  pages = {712--719},
  issn = {23521465},
  doi = {10.1016/j.trpro.2025.10.091},
  url = {https://linkinghub.elsevier.com/retrieve/pii/S2352146525007525},
  }

@incollection{jordanIntroductionVariationalMethods1998,
  title = {An {{Introduction}} to {{Variational Methods}} for {{Graphical Models}}},
  booktitle = {Learning in {{Graphical Models}}},
  author = {Jordan, Michael I. and Ghahramani, Zoubin and Jaakkola, Tommi S. and Saul, Lawrence K.},
  editor = {Jordan, Michael I.},
  date = {1998},
  pages = {105--161},
  publisher = {Springer Netherlands},
  location = {Dordrecht},
  doi = {10.1007/978-94-011-5014-9_5},
  url = {http://link.springer.com/10.1007/978-94-011-5014-9_5},
  }

@article{leeGibbsSamplerCoordinate2022,
  title = {Gibbs Sampler and Coordinate Ascent Variational Inference: {{A}} Set-Theoretical Review},
  shorttitle = {Gibbs Sampler and Coordinate Ascent Variational Inference},
  author = {Lee, Se Yoon},
  date = {2022-03-19},
  journaltitle = {Communications in Statistics - Theory and Methods},
  volume = {51},
  number = {6},
  pages = {1549--1568},
  issn = {0361-0926, 1532-415X},
  doi = {10.1080/03610926.2021.1921214},
  url = {https://www.tandfonline.com/doi/full/10.1080/03610926.2021.1921214},
  }

@article{liSEASONALANALYSISFACTORS2017,
  title = {A {{SEASONAL ANALYSIS ON FACTORS AFFECTING BIKE-SHARING CHOICE}}: {{WITH A FOCUS ON AIR POLLUTION}}’{{S IMPACT}}},
  author = {Li, Weibo and Kamargianni, Maria},
  date = {2017},
  }

@article{liuInvestigatingUserPreferences2025,
  title = {Investigating User Preferences for Dockless Bike- and Electric Bike-Sharing through Tracking Usage Patterns},
  author = {Liu, Yang and Li, Liqiong and Liu, Kai and He, Mingwei and Shi, Zhuangbin},
  date = {2025-08},
  journaltitle = {Transport Policy},
  volume = {169},
  pages = {41--55},
  issn = {0967070X},
  doi = {10.1016/j.tranpol.2025.04.025},
  url = {https://linkinghub.elsevier.com/retrieve/pii/S0967070X25001672},
  }

@article{liuSharedMicromobilityMeets2025,
  title = {Shared Micro-Mobility Meets Bus: {{A}} Spatiotemporal Heterogeneity Analysis in {{Chinese}} Medium-Sized Cities},
  shorttitle = {Shared Micro-Mobility Meets Bus},
  author = {Liu, Xize and Chen, Jingxu and Chen, Xuewu and Ning, Jiang},
  date = {2025-12},
  journaltitle = {Transport Policy},
  volume = {174},
  pages = {103853},
  issn = {0967070X},
  doi = {10.1016/j.tranpol.2025.103853},
  url = {https://linkinghub.elsevier.com/retrieve/pii/S0967070X25003968},
  }

@article{mckenzieSpatiotemporalComparativeAnalysis2019,
  title = {Spatiotemporal Comparative Analysis of Scooter-Share and Bike-Share Usage Patterns in {{Washington}}, {{D}}.{{C}}.},
  author = {McKenzie, Grant},
  date = {2019-06},
  journaltitle = {Journal of Transport Geography},
  volume = {78},
  pages = {19--28},
  issn = {09666923},
  doi = {10.1016/j.jtrangeo.2019.05.007},
  url = {https://linkinghub.elsevier.com/retrieve/pii/S0966692319302741},
  }

@book{mclachlanFiniteMixtureModels2000,
  title = {Finite {{Mixture Models}}},
  author = {McLachlan, Geoffrey and Peel, David},
  date = {2000-09-18},
  series = {Wiley {{Series}} in {{Probability}} and {{Statistics}}},
  edition = {1},
  publisher = {Wiley},
  doi = {10.1002/0471721182},
  url = {https://onlinelibrary.wiley.com/doi/book/10.1002/0471721182},
 }

@article{mohiuddinExaminingMarketSegmentation2024,
  title = {Examining Market Segmentation to Increase Bike-Share Use and Enhance Equity: {{The}} Case of the Greater {{Sacramento}} Region},
  shorttitle = {Examining Market Segmentation to Increase Bike-Share Use and Enhance Equity},
  author = {Mohiuddin, Hossain and Fitch-Polse, Dillon T. and Handy, Susan L.},
  date = {2024-01},
  journaltitle = {Transport Policy},
  volume = {145},
  pages = {279--290},
  issn = {0967070X},
  doi = {10.1016/j.tranpol.2023.10.021},
  url = {https://linkinghub.elsevier.com/retrieve/pii/S0967070X23002950},
  }

@book{murphyMachineLearningProbabilistic2013a,
  title = {Machine Learning: A Probabilistic Perspective},
  shorttitle = {Machine Learning},
  author = {Murphy, Kevin P.},
  date = {2013},
  series = {Adaptive Computation and Machine Learning Series},
  edition = {4. print. (fixed many typos)},
  publisher = {MIT Press},
  location = {Cambridge, Mass.},
  isbn = {978-0-262-01802-9},
  }

@inproceedings{pobudzeiUserSegmentationBased2024,
  title = {User {{Segmentation}} Based on {{Usage Frequency}}: {{A Case Study}} of a {{Multimodal Shared Micromobility}} in a {{Non-Urban Campus Environment}}{\textsuperscript{*}}},
  shorttitle = {User {{Segmentation}} Based on {{Usage Frequency}}},
  booktitle = {2024 {{IEEE Intelligent Vehicles Symposium}} ({{IV}})},
  author = {Pobudzei, Maryna and Hoffmann, Silja},
  date = {2024-06-02},
  pages = {992--999},
  publisher = {IEEE},
  location = {Jeju Island, Korea, Republic of},
  doi = {10.1109/IV55156.2024.10588699},
  url = {https://ieeexplore.ieee.org/document/10588699/},
  }

@article{roig-costaUnderstandingMultimodalMobility2026,
  title = {Understanding Multimodal Mobility Patterns of Micromobility Users in Urban Environments: Insights from {{Barcelona}}},
  shorttitle = {Understanding Multimodal Mobility Patterns of Micromobility Users in Urban Environments},
  author = {Roig-Costa, Oriol and Marquet, Oriol and Arranz-López, Aldo and Miralles-Guasch, Carme and Van Acker, Veronique},
  date = {2026-06},
  journaltitle = {Transportation},
  volume = {53},
  number = {3},
  pages = {1335--1363},
  issn = {0049-4488, 1572-9435},
  doi = {10.1007/s11116-024-10531-3},
  url = {https://link.springer.com/10.1007/s11116-024-10531-3},
  }

@article{samadzadWhatAreFactors2023,
  title = {What Are the Factors Affecting the Adoption and Use of Electric Scooter Sharing Systems from the End User's Perspective?},
  author = {Samadzad, Mahdi and Nosratzadeh, Hossein and Karami, Hossein and Karami, Ali},
  date = {2023-06},
  journaltitle = {Transport Policy},
  volume = {136},
  pages = {70--82},
  issn = {0967070X},
  doi = {10.1016/j.tranpol.2023.03.006},
  url = {https://linkinghub.elsevier.com/retrieve/pii/S0967070X23000598},
  }

@article{talavera-garciaExaminingSpatiotemporalMobility2021,
  title = {Examining Spatio-Temporal Mobility Patterns of Bike-Sharing Systems: The Case of {{BiciMAD}} ({{Madrid}})},
  shorttitle = {Examining Spatio-Temporal Mobility Patterns of Bike-Sharing Systems},
  author = {Talavera-Garcia, Ruben and Romanillos, Gustavo and Arias-Molinares, Daniela},
  date = {2021-01-01},
  journaltitle = {Journal of Maps},
  volume = {17},
  number = {1},
  pages = {7--13},
  issn = {1744-5647},
  doi = {10.1080/17445647.2020.1866697},
  url = {https://www.tandfonline.com/doi/full/10.1080/17445647.2020.1866697},
  }

@article{veveEstimationSharedMobility2020,
  title = {Estimation of the Shared Mobility Demand Based on the Daily Regularity of the Urban Mobility and the Similarity of Individual Trips},
  author = {Veve, Cyril and Chiabaut, Nicolas},
  editor = {Kato, Hironori},
  date = {2020-09-17},
  journaltitle = {PLoS ONE},
  volume = {15},
  number = {9},
  pages = {e0238143},
  issn = {1932-6203},
  doi = {10.1371/journal.pone.0238143},
  url = {https://dx.plos.org/10.1371/journal.pone.0238143},
  }

@article{wainwrightGraphicalModelsExponential2008,
  title = {Graphical {{Models}}, {{Exponential Families}}, and {{Variational Inference}}},
  author = {Wainwright, Martin J. and Jordan, Michael I.},
  date = {2008-12-18},
  journaltitle = {Foundations and Trends® in Machine Learning},
  volume = {1},
  number = {1--2},
  pages = {1--305},
  issn = {1935-8237, 1935-8245},
  doi = {10.1561/2200000001},
  url = {https://www.emerald.com/ftmal/article/1/1-2/1/1332382/Graphical-Models-Exponential-Families-and},
  }

@article{wintersWhoAreSuperusers2019,
  title = {Who Are the ‘Super-Users’ of Public Bike Share? {{An}} Analysis of Public Bike Share Members in {{Vancouver}}, {{BC}}},
  shorttitle = {Who Are the ‘Super-Users’ of Public Bike Share?},
  author = {Winters, Meghan and Hosford, Kate and Javaheri, Sana},
  date = {2019-09},
  journaltitle = {Preventive Medicine Reports},
  volume = {15},
  pages = {100946},
  issn = {22113355},
  doi = {10.1016/j.pmedr.2019.100946},
  url = {https://linkinghub.elsevier.com/retrieve/pii/S2211335519301202},
  }

@article{padghamDodgrPackageNetwork2019,
  title = {Dodgr: {{An R}} Package for Network Flow Aggregation},
  shorttitle = {Dodgr},
  author = {Padgham, Mark},
  date = {2019-01-14},
  journaltitle = {Transport Findings},
  doi = {10.32866/6945},
  url = {https://findingspress.org/article/6945-dodgr-an-r-package-for-network-flow-aggregation},
}

@article{fruchterman1991graph,
  title={Graph drawing by force-directed placement},
  author={Fruchterman, Thomas M. J. and Reingold, Edward M.},
  journal={Software: Practice and Experience},
  volume={21},
  number={11},
  pages={1129--1164},
  year={1991},
  publisher={Wiley},
  doi={10.1002/spe.4380211102}
}

\appendix

\section{Simulation Study}
\label{sec:simulation}
\subsection{Data design}
A simulation study was conducted to evaluate the performance of the proposed variational Bayes product-multinomial framework to recover latent user groups from repeated categorical trip records under controlled conditions.

A synthetic mobility data set comprising 600 users is generated that is distributed in 8 latent clusters. Each cluster was characterized by distinct behavioral profiles defined by origin and destination locations, vehicle-pass usage, seasonal travel patterns, and trip-duration distributions. For each user, the number of trips was generated from a cluster-specific Poisson distribution, resulting in varying levels of activity between users. Individual trips were then generated according to cluster-specific multinomial probability distributions for the same set of factors defined above. 

The simulated data design is used to depict the same nature as the hierarchical structure of the observed data, where multiple trips are nested within users and user behavior is represented through repeated categorical observations.

The variational Bayes product-multinomial model is fitted to the simulated data set using the procedure described in Section 4. The model selection is performed using ELBO for K values ranging from 5 to 12 clusters. The performance of the model is assessed by comparing the estimated cluster assignments with the true simulated membership using the Adjusted Rand Index (ARI).

\subsection{Results}
The ELBO reached its maximum value at \(K=8\) (Figure ~\ref{fig:appendix_elbo}), corresponding to the true number of latent clusters used in the data generation process. The resulting ARI was 0.939, indicating a strong agreement between the estimated and true clustering structures. Overall, 597 of 650 simulated users (91.8\%) are correctly assigned to their generating cluster (Table ~\ref{tab:simulation}). 
The clusters obtained show an interesting behavior pattern based on the simulated data. According to classified zones, the clusters reveal localized users, tourists, peripheral users, and others with high precision. 

\begin{table}[H]
\centering

\label{tab:simulation}
\begin{tabular}{lc}
\hline
Metric & Value \\
\hline
True number of clusters & 8 \\
Selected number of clusters & 8 \\
Adjusted Rand Index (ARI) & 0.939 \\
Users correctly classified & 597 / 650 \\
Classification accuracy (\%) & 91.8 \\
Misclassified users & 53 \\
\hline

\end{tabular}
\caption{Performance of the variational Bayes clustering algorithm on simulated mobility data.}
\end{table}

\begin{figure}[H]
    \centering
    \includegraphics[width=1.0\linewidth]{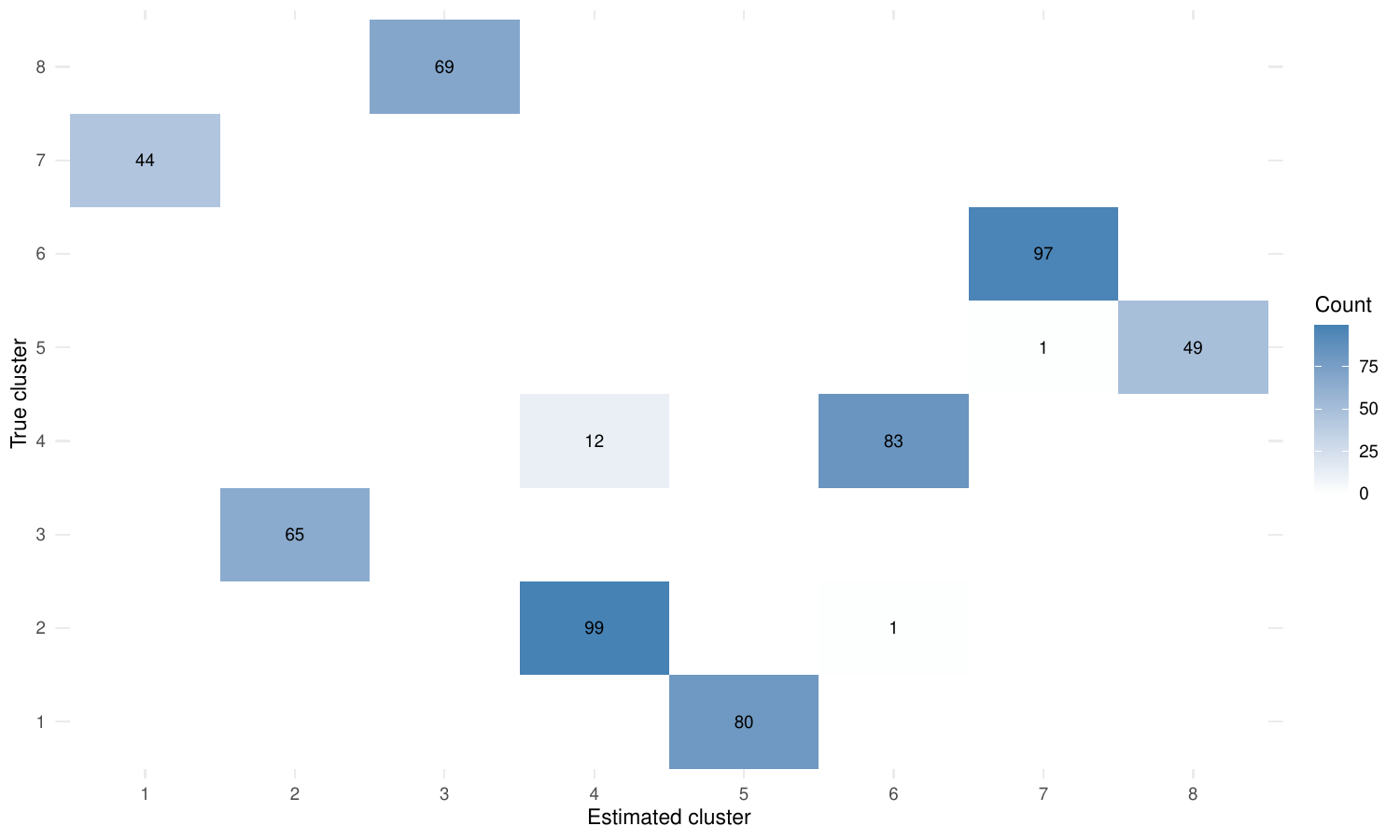}
    \caption{Confusion matrix for simulated results}
    \label{fig:placeholder}
\end{figure}

\begin{figure}[H]
    \centering
    \includegraphics[width=1.0\linewidth]{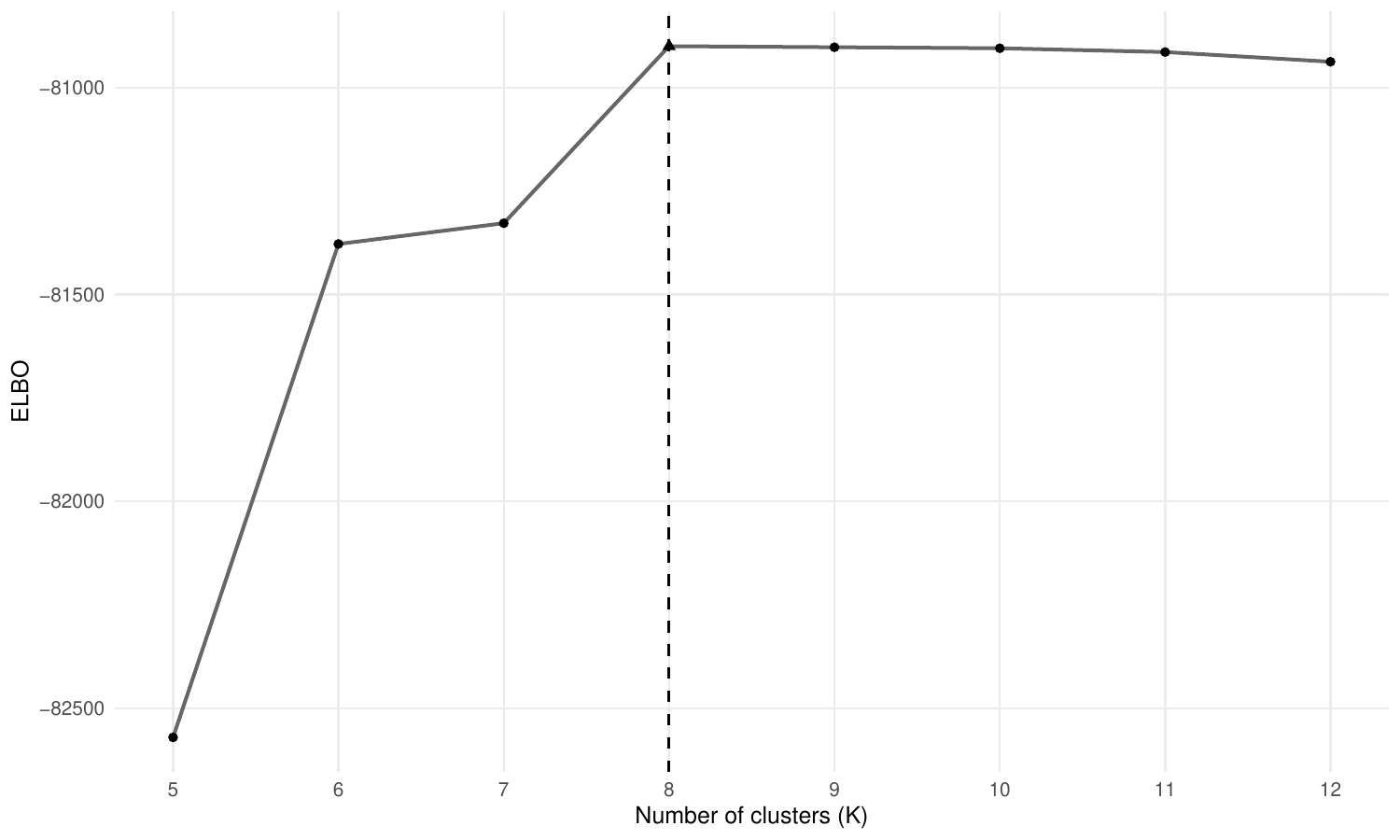}
    \caption{Model selection from simulated data}
    \label{fig:appendix_elbo}
\end{figure}

The confusion matrix comparing the true and estimated membership shows recovery for almost all of the simulated clusters (Figure ~\ref{fig:placeholder}). The misclassification was limited to groups with similar spatial structures, indicating the difficulty in classifying highly related mobility behavior. In general, the results show a high ARI value with strong recovery. These results indicate that the proposed framework is capable of recovering a known latent clustering structure from repeated categorical trip records. 

\end{document}